\documentclass[prd,aps,twocolumn]{revtex4-2}
\usepackage{graphicx} 
\usepackage{amsmath,amssymb}
\usepackage{aas_macros}
\usepackage{color}
\begin{document}
\title{{\it Gaia} 400,894 QSO constraint on the energy density of\\ 
low-frequency gravitational waves}
\author{Shohei Aoyama}
\email[]{aoyama@icrr.u-tokyo.ac.jp}
%\altaffiliation{}
\affiliation{Institute for Cosmic Ray Research, The University of Tokyo, Kashiwanoha 5-1-5, Kashiwa, Chiba 277-8582, Japan}
\author{Daisuke Yamauchi}
\email[]{yamauchi@jindai.jp}
%\altaffiliation{}
\affiliation{Faculty of Engineering, Kanagawa University, Kanagawa, 221-8686, Japan}
\author{Maresuke Shiraishi}
\email[]{shiraishi-m@t.kagawa-nct.ac.jp}
%\altaffiliation{}
\affiliation{Department of General Education, National Institute of Technology, Kagawa College, 355 Chokushi-cho, Takamatsu, Kagawa 761-8058, Japan}
\author{Masami Ouchi}
\email[]{ouchims@icrr.u-tokyo.ac.jp}
\affiliation{National Astronomical Observatory of Japan, Osawa 2-21-1, Mitaka, Tokyo 181-8588, Japan}
\affiliation{Institute for Cosmic Ray Research, The University of Tokyo, Kashiwanoha 5-1-5, Kashiwa, Chiba 277-8582, Japan}
\affiliation{Kavli Institute for the Physics and Mathematics of the Universe (Kavli IPMU, WPI), The University of Tokyo, 
Kashiwanoha 5-1-5, Kashiwa, Chiba 277-8583, Japan}
\date{\today}

\begin{abstract}% 257 words / criteria=[ < 500 words & about 5% of article length ]
	%長周期重力波は光を曲げてapparent proper motions of QSOを引き起こすこと。
	%先行研究ではVLBIが使われ、711天体が観測された。観測精度は非常に良い(0.2 mas)。しかしVLBIは天体数、観測時間いずれも伸ばすことは現行の装置では難しいこと。
	%Gaiaでは400,894 QSOsを同じオーダーの精度で観測していること。
	%結果としてほぼ二桁向上した精度の制限を得たこと。

Low frequency gravitational waves (GWs) are keys to understanding cosmological inflation and super massive blackhole (SMBH) formation via blackhole mergers, while it is difficult to identify the low frequency GWs with ground-based GW experiments such as the advanced LIGO (aLIGO) and VIRGO due to the seismic noise. Although quasi-stellar object (QSO) proper motions produced by the low frequency GWs are measured by pioneering studies of very long baseline interferometry (VLBI) observations with good positional accuracy, the low frequency GWs are not strongly constrained by the small statistics with 711 QSOs (Darling et al. 2018). Here we present the proper motion field map of 400,894 QSOs of the Sloan Digital Sky Survey (SDSS) with optical {\it Gaia} EDR3 proper motion measurements whose positional accuracy is $< 0.4$ milli-arcsec comparable with the one of the radio VLBI observations. We obtain the best-fit spherical harmonics with the typical field strength of $\mathcal{O}(0.1)\, \mu$arcsec, 
and place a tight constraint on the energy density of GWs, 
%$\Omega_{\rm gw}< 0.4768 \times 10^{-3}$ (95 \% confidence level), 
$\Omega_{\rm gw}=(0.964 \pm 3.804) \times 10^{-4}$ (95 \% confidence level), 
that is significantly stronger than the one of the previous VLBI study by two orders of magnitude at the low frequency regime of $f <10^{-9}\,{\rm [Hz]}\simeq (30\,{\rm yr})^{-1}$ unexplored by the pulsar timing technique. Our upper limit rules out the existence of SMBH binary systems at the distance 
$r < 400$ kpc from the Earth where the Milky Way center and 
local group galaxies are included. 
Demonstrating the limit given by our optical QSO study, we claim that astrometric satellite data including the forthcoming {\it Gaia} DR5 data with small systematic errors are powerful to constrain low frequency GWs.
%(257 words)
\end{abstract}

\maketitle

\section{Introduction}
%重力波の一般的な導入(長所、予言、発見)
%重力波は一般相対性理論の最も重要な予言の1つである。
%重力波は視線上の物質とほとんど相互作用しないのでblackhole mergerや宇宙論的インフレーションといった強い重力場の性質を直接観測できる。
%一方で振幅は非常に小さいので
%重力波の観測を困難にしてきた。
Existence of gravitational waves (GWs) is one of the primary predictions of general relativity. Due to little interaction with matter on the line of sight, one can directly investigate the nature of sources in the strong gravitational fields such as mergers of binaries of blackholes and the cosmological inflation. However, because
the strain amplitude of GWs is expected to be very small $\lesssim 10^{-24}$, there are many difficulties in the direct detection of GWs.

%重力波の直接検測には約1世紀を要した。
%2015年に初めてレーザー干渉系で重力波の観測に成功した。この観測でブラックホール連星系の存在が初めて観測的に確認された。
%さらに2017年に中性子星連星系からの重力波の観測に成功した。
%この重力波イベントでは電磁波対応天体の観測に成功した。
%電磁波対応天体が重力波の信号の1.7秒後に観測されたことから
%重力波と電磁波は同じ速さで進むことが示された。
%これにより2つの代表的な理論が棄却された。
%GWの観測でハッブル定数も測定された。
%10回の測定で3\%の制度になる。
About one century has been past to detect GWs directly since the prediction about the existence of GWs.
In 2015, GWs are directly detected for the first time with a laser interferometer named the advanced LIGO (aLIGO) \cite{2016PhRvL.116f1102A}. With the event named GW150914, the existences of GWs as well as a binary blackhole merger have been observationally confirmed. In addition, in 2017, GWs from a binary neutron star (NS) merger are detected with aLIGO and Virgo. This binary NS merger is named GW170817. Simultaneously, the optical counterpart of GW170817, GRB170817A (AT2017gfo), is detected at a wide frequency range from gamma-ray to radio waves. Because the gamma ray from GRB170817A is detected only 1.7 seconds after the GW-signal detections, it is confirmed that the speed of GWs $c_{\rm gw}$ is the same as the speed of light $c$ with $10^{-15}$ accuracy \cite{2017ApJ...848L..13A}. These observations reject the modified gravity theories which predict $c_{\rm gw}\ne c$ \cite{2018PhRvD..97f1501L,2019PhRvD.100f3509P}. As a result, for example, the covariant Galileon (e.g. \cite{2009PhRvD..79h4003D}) and the Gauss-Bonnet gravity (see \cite{2003inco.book.....R}) are ruled out, 
and one needs an alternative theoretical framework for 
the accelerating expansion of the current Universe. The event GW170817/GRB170817A also provides a new standard ruler to astronomy 
because the absolute luminosity of the NS binary merger is estimated by numerical relativity and hydrodynamical simulations with the effects of general relativity. Due to the nature of the Hubble-Lema\^{i}tre law, the standard ruler can be used to estimate the Hubble constant. The Hubble constant is successfully measured with GW170817/GRB170817A by these studies \cite{2017Natur.551...85A, 2019NatAs...3..940H}. These studies prove that GW astronomy provides a new and powerful ruler in the Universe. By measuring more than 10 NS binary mergers in the future, the Hubble constant will be determined at the level of 3\% accuracy \cite{1986Natur.323..310S,2019NatAs...3..940H}, which is independent from measurements given by type Ia supernova projects. 
Observations of GWs provide precise cosmology, and reveal mechanisms of structure formation in the Universe.

%重力波観測の現状、スペクトラムの存在、長周期重力波の重要性。
%%①現在の重力波干渉計は30Hz~500Hzにしか感度がない。これは電磁波の波長域に比べれば
%%非常に狭い。
%%②長周期重力波は中間質量ブラックホール連星系の特定や
%%大質量ブラックホール形成モデルやインフレーション模型の区別に使うことができる。
%%③それに加えて、宇宙ひもなどの位相欠陥やアクシオンなどの超低質量粒子からの重力波も
%%この振動数帯に入る。
%%④よってこの振動数帯の重力波の観測感度の向上は天文学非常に重要である。
Although GWs are powerful probes of the Universe, the current observable frequency range of GWs (30Hz $\sim $ 500 Hz) is much narrower than that of electromagnetic waves  ($10^{8}$ Hz $\sim $ $10^{26}$ Hz). 
Especially, upper limits of the energy density of GWs at the low frequency range are important to identify the binaries of intermediate-mass blackholes. The low frequency GWs can distinguish the responsible models of the cosmological inflation and the formation of super massive blackholes (SMBHs) from mergers of intermediate-mass blackholes. 
In addition, the low frequency GWs provide critical tests on a number of models of topological defects such as cosmic strings and the ultra-light pseudo-scalar fields such as axions. However, aLIGO and Virgo cannot detect GWs at the low frequency range due to the seismic noise. In order to overcome the difficulty of the seismic noise, space-based interferometers have been proposed. The laser interferometer space antenna (LISA) and Deci-hertz interferometer gravitational wave observatory (DECIGO) will have sensitivity at the frequency regimes, $10^{-3}\,{\rm Hz}$ and $10^{-1}\,{\rm Hz}$, respectively. LISA can detect binary mergers of intermediate-mass blackholes with a total mass of $10^{3}\,{\rm M}_{\odot}$ mostly over the observable Universe. In addition, the GWs from cosmological inflation can be directly identified with DECIGO.

%長周期重力波の制限の困難。
%% 1. aLIGO/Virgoなどの重力波干渉計では重力波を波動として捉えている。
%% 2. 長周期重力波は周期が観測装置の寿命を越えていることから
%%%   単一の観測装置では波として観測するのは難しい。
%% 3. 長周期重力波には天球面上の天体に見かけ上の固有運動を発生させる性質がある。
%% 4. intrinsicな固有運動が無視できる銀河系外の点源の見かけ上の固有運動を測れば
%% 5. この見かけ上の固有運動を測定することにより、長周期重力波を制限できる。
%% 6. 今までには測定精度が良いVLBIが使われてきた。
%% 7. VLBIでは現在までに711天体の
%%%   proper motionを測定し、エネルギー密度に制限をつけた。
%% 8. VLBIには2つの大きな問題がある。
%% 9. VLBIでは明るい天体しか観測できない。
%%10. VLBIでは同時に複数の望遠鏡を長時間にわたって占有しなければならない。
%
The experiments which have succeeded in detecting GWs identify GWs as waves.
There is a difficulty to detect GWs at the low frequency range $f<10^{-9}\,{\rm Hz}$ as waves with current experimental facilities. In order to detect these GWs as waves, these facilities need to be operated more than 30 years, which is longer than the typical lifetime of the experimental facilities. Thus alternative methods for detecting low frequency GWs are needed.
A number of researchers have found that low frequency GWs induce the apparent proper motions of point sources because GWs bend the path of photons regardless of the frequency of GWs. Because the intrinsic proper motions of extragalactic point sources are negligibly small, the quasi-stellar objects (QSOs) become favorable targets.  Because the apparent proper motion induced by GWs is estimated to be very small ($\ll 1\,\mu$arcsec), a large amount of precise observational data are required. In radio observations, the very long baseline interferometry (VLBI) technique can be used. With VLBI, $<0.2$-milli-arcsecond (mas) precision has been achieved. A pioneering work of this technique is Gwinn et al. (1997) \cite{1997ApJ...485...87G}, hereafter G97. Darling et al. (2018) \cite{2018ApJ...861..113D}, hereafter D18, measure proper motions of 711 objects with the VLBI and set a upper limit on low frequency GWs, $\Omega_{\rm gw} < 0.64\times 10^{-2} $ for $f<10^{-9}$ Hz. 
While the VLBI technique is successful, this technique has two major limitations. 
First, the targets of VLBI are limited to radio loud objects necessary for VLBI detections. Thus it is difficult to increase the number of the targets. 
Second, these VLBI studies need a number of radio antennae for a long time $\sim \mathcal{O}(1)\,{\rm yr}$ for interferometric observations. Thus it is difficult to improve the sensitivity of very low frequency GWs with the existing VLBIs. 

%MO: Because there are too many revisions, I won't add comments anymore.
Astrometry in optical wavelengths are also useful to measure low frequency GWs \cite{1997ApJ...485...87G,2018ApJ...861..113D}. One of advantages of optical observations is that one can measure positions of astronomical objects accurately enough for low frequency GW detections with a single telescope by no interferometric technique. In addition, due to a large number of observable targets in optical wavelengths, it is easy to increase the number of targets. Because, in optical wavelengths, the atmosphere makes undesirable fluctuations of the positions of targets, space-borne observations are required for precise astrometry. {\it Gaia} is a satellite for astrometry, measuring proper motions of stars in the Milky Way and extragalactic objects including QSOs in the optical wavelength \cite{2016A&A...595A...1G}. {\it Gaia} has measured proper motions of extragalactic sources brighter than $g$ band magnitude $G<22$. 
{\it Gaia} early data release 3 (EDR3) has archived the proper motion accuracy $\sim$ 400 $\mu$arcsec/yr at $G\simeq 20$ for an individual object that is accomplished by the post processes \cite{2020arXiv201201533G}. Because the estimation methods of proper motions from raw data are improved, the systematic errors of proper motions in EDR3 are reduced typically by three times better than those in DR2 \cite{2020arXiv201203380L}. Strong constraints on GWs can be placed with the EDR3 data. Several authors (e.g. \cite{1999BaltA...8..239C, 2017PhRvL.119z1102M, 2018ApJ...861..113D, 2018CQGra..35d5005K, 2020arXiv201002218W}) have already claimed that {\it Gaia} is a very powerful tool to study GWs. However, no constraints on the energy density of GWs at very low frequencies ($f<10^{-9}\,{\rm [Hz]}$) have been reported yet, probably due to the difficulty of a large QSO sample development (Section \ref{sample}) and the moderately large systematic errors of the previous {\it Gaia} data of DR1 and DR2.

In this paper, we place an upper limit on the energy density of low frequency GW with {\it Gaia} EDR3 data, adapting the formula claimed by Mignard \& Klioner (2012) \cite{2012A&A...547A..59M}, hereafter referred to as MK12. We fit the proper motions measured by {\it Gaia} with the formula of vector harmonics. Performing the vector fitting, we study a signal of low-frequency GWs.

This paper is organized as follows. In section \ref{Sec_method}, we introduce the methods of this study. Section \ref{sample} describes the construction procedure of QSO samples with the Sloan Digital Sky Survey (SDSS) data. In section \ref{calculation_results}, we show the upper limit of the very low frequency GWs at $f< 10^{-9}$ Hz. In section \ref{discussion}, we discuss the detectability of inspiral phases of SMBH binaries, and describe future prospects for a GW constraint with the forthcoming {\it Gaia} DR5 data. Section \ref{conclusion} concludes the paper.

\section{Method}\label{Sec_method}
	%最低限書かなければならないこと
%①解析手法(vector harmonics、アルゴリズムlmfit)
%段落構成
 	%第1段落　QSOのproper motion -> 観測範囲で重力波以外では0になることの説明
%第２段落　サンプルに対するvector hermonicsの適用->エネルギー密度$\Omega(f<10^{-9}Hz)$の上限値得かた
		%第３段落　誤差の計算方法(山内さんのノート)
	%\subsection{Vector harmonics for GW energy density}
The low frequency GWs such as $f<10^{-9}\,[{\rm Hz}]\simeq (30\,{\rm yr})^{-1}$ are detected as apparent proper motions of extragalactic point sources such as QSOs (Section 1). The GWs create quadrupole modes ($\ell =2$) of the proper motions 
\footnote{Strictly speaking, GWs creates multipole modes $(\ell \ge 2)$ (see \cite{2019PhRvD.100b1303N}). However, because the quadruple mode ($\ell =2$) is dominated in the generated proper motions with low frequency GWs, we only focus on the quadruple mode of proper motions. }. Because the intrinsic proper motions of QSOs are negligible due to a large separation between QSOs and the observer, the proper motion of QSOs can be used to constrain the amplitude of GWs. We aim at detecting low frequency GWs with QSOs.

In order to identify the quadrupole proper motion, we perform the vector harmonics analysis suggested by MK12. A position on the spherical coordinate system $(\theta, \varphi)$ is related with the one of the equatorial coordinate system ($\alpha, \delta$) by the following equations: 
\begin{eqnarray}
\alpha &=&\varphi \, \\
\delta &=&\dfrac{1}{2}\pi -\theta\, . 
\end{eqnarray}
One can define the unit vectors on the spherical coordinate system as $(\vec{e}_{\theta},\vec{e}_{\varphi})$. The unit vectors yield  
\begin{eqnarray}
\vec{e}_{\theta}\cdot \vec{e}_{\theta}&=&1\, ,\label{nob1}\\
\vec{e}_{\varphi}\cdot \vec{e}_{\varphi}&=&1\, ,\label{nob2}\\
\vec{e}_{\theta}\cdot \vec{e}_{\varphi}&=&0\, . \label{nob3}
\end{eqnarray}
With equations \eqref{nob1} -- \eqref{nob3}, the proper motion fields $\vec{\mu}$ can be uniquely decomposed as
\begin{eqnarray}
\vec{\mu}=\mu_{\theta}\vec{e}_{\theta}+\mu_{\varphi}\vec{e}_{\varphi}\, .
\end{eqnarray}
Mignard \& Klioner (2012) proposes the mathematical formula of the multipole decomposition of  $\vec{\mu}$ as 
\begin{equation}
	\vec{\mu}(\theta, \varphi) = \displaystyle\sum_{\ell =1}^{\infty}\sum_{m=-\ell}^{+\ell}\left(
E_{\ell m}\vec{Y}^{\rm E}_{\ell m}(\theta, \varphi)+
B_{\ell m}\vec{Y}^{\rm B}_{\ell m}(\theta, \varphi)
\right)~,\label{decompose1}
\end{equation}
where $\vec{Y}^{\rm E}_{\ell m}(\theta, \varphi)$ and $\vec{Y}^{\rm B}_{\ell m}(\theta, \varphi)$ are E-mode and B-mode eigenvectors, respectively (see Appendix A). $E_{\ell m}$ and $B_{\ell m}$ are the amplitudes of the proper motion fields. Note that $E_{\ell m}$ and $B_{\ell m}$ are complex sequences for $\ell $ and $m$. The energy density of GWs $\Omega_{\rm gw}$ at the range $f<10^{-9}\, {\rm Hz}$ is \cite{2018ApJ...861..113D}
\begin{equation}
\Omega_{\rm gw}=\dfrac{3}{2\pi \tilde{H}_{0}^2}
\sum_{m=-2}^{2}\left(
|E_{2,m}|^{2}+|B_{2,m}|^{2}
\right)~,\label{omegaGW}
\end{equation}
where $\tilde{H}_{0}$ is the Hubble constant in units of yr$^{-1}$ that is $\tilde{H}_{0}=14.20\,\mu{\rm arcsec}\slash{\rm yr}$ (Planck collaboration 2018 \cite{2020A&A...641A...6P}). 
$E_{2, m}$ and $B_{2, m}$ are the values of $E_{\ell m}$ and $ B_{\ell m}$ at $\ell = 2$, respectively. The energy density of GWs can be also written with the strain amplitude 
$h_{\rm c}$ as 
\begin{equation}
\Omega_{\rm gw}=\dfrac{10\pi^{2}}{3H_{0}^{2}}\left(fh_{\rm c} \right)^{2}, \label{OmegaGW_hc}
\end{equation}
where $H_{0}$ is the Hubble constant in units of s$^{-1}$ that is $H_{0}=2.1830\times 10^{-18}[{\rm s}^{-1}]$ (Planck collaboration 2018 \cite{2020A&A...641A...6P}).
Equation \eqref{OmegaGW_hc} is equivalent with
\begin{equation}
h_{\rm c}=\dfrac{H_{0}}{\pi f}\sqrt{\dfrac{3}{10}\Omega_{\rm gw}}\label{hc_OmegaGW}
\end{equation}

We compare the proper motion field of a model $\vec{\mu}^{\rm th}$ and observational data $\vec{\mu}^{\rm obs}$.  Because of the orthonormality of the unit vectors described in equation \eqref{nob1}, $\vec{\mu}^{\rm th}$ can be written as
\begin{eqnarray}
\mu_{\theta}^{\rm th} &=& \displaystyle\sum_{\ell =1}^{\infty}
\sum_{m=-\ell}^{+\ell}\left(
E_{\ell m}\vec{e}_{\theta}\cdot \vec{Y}^{\rm E}_{\ell m}+
B_{\ell m}\vec{e}_{\theta}\cdot \vec{Y}^{\rm B}_{\ell m} \right)~,\\
\mu_{\varphi}^{\rm th} &=& \displaystyle\sum_{\ell =1}^{\infty}
\sum_{m=-\ell}^{+\ell}\left(
E_{\ell m}\vec{e}_{\varphi}\cdot \vec{Y}^{\rm E}_{\ell m}+
B_{\ell m}\vec{e}_{\varphi}\cdot \vec{Y}^{\rm B}_{\ell m}\right)~.
\end{eqnarray}
Appendix gives the coefficients 
$\vec{e}_{\theta}\cdot \vec{Y}^{\rm E}_{\ell m}$,  
$\vec{e}_{\varphi}\cdot \vec{Y}^{\rm E}_{\ell m}$. 
$\vec{e}_{\theta}\cdot \vec{Y}^{\rm B}_{\ell m}$,  
$\vec{e}_{\varphi}\cdot \vec{Y}^{\rm B}_{\ell m}$. 

We estimate $E_{\ell m}$ and $B_{\ell m}$ by fitting 
$\vec{\mu}^{\rm obs}$ to $\vec{\mu}^{\rm th}$. 
We use a least-square approach that is proposed in \cite{Data_analysis2006}. 
In this approach, we define the positive quantity $r$ as 
\begin{equation}
r \equiv
\underset{E_{\ell m}, B_{\ell m}}{\min}
\left[
\displaystyle\sum_{k=1}^{N_{\rm sample}}
\left\{ 
\dfrac{
\left(
\mu^{\rm obs}_{\theta, k}-\mu^{\rm th}_{\theta}
\right)^{2}
}{ 
(\sigma_{\theta, k}^{\rm obs} )^{2} 
} 
+
\dfrac{
\left(
\mu^{\rm obs}_{\varphi, k}-\mu^{\rm th}_{\varphi}
\right)^{2}
}{ 
(\sigma_{\varphi, k}^{\rm obs} )^{2} 
}
\right\}
\right]\, ,
\end{equation}
where the index $k$ is a number of the ID of the sample. The value of  
$\sigma_{\theta, k}^{\rm obs}$
$(\sigma_{\varphi, k}^{\rm obs})$ is the error of the proper motion for the direction $\theta (\varphi )$ of the $k$-th QSO in the sample. 
$N_{\rm sample}$ is the size of SDSS-{\it Gaia} sample, which is 400,894.
In order to obtain $E_{\ell m}$ and $B_{\ell m}$, we use a python package {\tt lmfit}. Note that both $E_{\ell m}$ and $B_{\ell m}$ are generally complex numbers. Because the fitted vectors should be real, 
$E_{\ell m}\vec{Y}_{\ell m}^{\rm E}$ and 
$B_{\ell m}\vec{Y}_{\ell m}^{\rm B}$ yield 
\begin{eqnarray}
E_{\ell m}\vec{Y}_{\ell m}^{\rm E}
&=&{\rm Re}(E_{\ell m}){\rm Re}(\vec{Y}_{\ell m}^{\rm E})
-{\rm Im}(E_{\ell m}){\rm Im}(\vec{Y}_{\ell m}^{\rm E})~,\notag \\
& & \\
B_{\ell m}\vec{Y}_{\ell m}^{\rm B}&=&
{\rm Re}(B_{\ell m}){\rm Re}(\vec{Y}_{\ell m}^{\rm B})
-{\rm Im}(B_{\ell m}){\rm Im}(\vec{Y}_{\ell m}^{\rm B})~,\notag \\
& &
\end{eqnarray}
where ${\rm Re}(X)$ and ${\rm Im}(X)$ are the real and imaginary part of $X$, respectively. In this paper, 
we consider the quadruple mode ($\ell = 2$) of proper motions which originate low frequency GWs. For the construction of the proper motion fields, we have 10 quantities to be fitted. Thus, the quantities are 
${\rm Re}(E_{2, m})$, 
${\rm Re}(B_{2, m})$, 
${\rm Im}(E_{2, m})$, 
${\rm Im}(B_{2, m})$, 
$E_{2, 0}$, and $B_{2, 0}$
at $m=1, 2$. Here $E_{\ell, 0}$, and $B_{\ell, 0}$ are real quantities. 
We adopt $[-0.3, 0.3](\mu {\rm arcsec})$ as prior ranges 
because the sum of each component is constrained by $\ell (\ell +1)\sum_{m=-2}^{2}(|E_{\ell m}|^{2}+|B_{\ell m}|^{2}|)< 1 \mu {\rm arcsec}^{2}$ by Darling et al (2018). We check the dependence of the prior range. If one adopts a prior range beyond $[-0.3, 0.3]$, 
the chi-square value of the fitting becomes un-physically large.

\section{Sample}\label{sample}
		%第１段落　SDSS QSO (spec) sample
		%第２段落　SDSS QSOとGAIAソースのcross match（KD-tree法を適用）→サンプル完成
%この下の階層の流れ( 一文につき1行)

%SDSS QSO sampleを使う。
%SDSS QSO sampleを作るときにはcontaminationを除くべき。
%contaminationを除くためにspec. Sampleを使う。
%その結果、0.8 million objects are identified.

%SDSS-QSO とGaiaの天体をcross-match
%2つのサンプルからのソースの距離が0.5 arcsec未満であるとき、同一とみなす。
%KD tree法でcross machした。
%10^4個見つかった。

%SDSS-QSO->Gaia counterpart : 違う天体を同一視する可能性がある。
%Gaiaの個数密度が小さい(~1/arcmin^{2})ので上記の可能性は無視しうる。
%cross-matchで得られた天体数は0.8 millionなので上記の期待値は8個になり無視しうる。

%\subsection{SDSS-{\it Gaia} QSO sample}
	We use the QSO sample constructed with spectroscopic data taken by 
the SDSS. 
Because the apparent proper motion originated from very long GWs is expected to be small, we should remove the contamination objects such as Galactic stars and galaxies. 
For this purpose, we choose to use spectroscopically confirmed QSOs 
in the 16th data release (DR16) of SDSS that is composed of 817,402 QSOs that are referred to as the SDSS-QSO sample.  
We cross-match objects of the SDSS-QSO sample with those of {\it Gaia} EDR3. In cross-matching between objects taken from the two samples, we regard two objects with a 
distance less than 0.5 arcsec as an identical object. 
%KD tree法でcross machした。
%cross-matchを普通にすると終わらない(the calculation cost)のでKD-tree methodを使います。
%0.4*10^6個見つかった。
%計算コストをセーブするためにKD-tree法でcross-matchする。
%The KD-tree method enables cross-match with the cost $\log_{2}(N)$ for each object. 
We adopt the KD-tree method 
\footnote{The KD-tree method enables to cross-match the objects with the cost $\log_{2}(N)$ for each object. Here $N$ is the size of the sample.}
as the algorithm of cross-matching in order to complete the cross-match within a reasonable time.
Finally, we find that proper motions of 400,894 QSOs out of 817,402 QSOs in the SDSS-QSO sample. We call 400,894 QSOs the SDSS-{\it Gaia} QSO sample. 
Exploiting the SDSS-{\it Gaia} sample of which we measure the proper motion of QSOs, we obtain a upper limit on the energy density of the very long GWs.

%However, 数密度が低い,→1個/ per 1 arcmin^2 [Gaia paper] (一文). 
%たまたま1個のQSOが違うobjectとcross-match 
%The expected mistakenly cross-matched number is 8 and it is small enough.

There is a possibility that the SDSS QSO counterpart can be a different object (i.e. star or galaxy) of the {\it Gaia} EDR3 sample that exists, by chance, in a $<0.5$-arcsec distance. 
However, the surface density of objects detected by {\it Gaia} is very small ($\lesssim 1\,$arcmin$^{-2}$). Because the size of SDSS-QSO sample is 817,402, the number of the mistakenly cross-matched objects is expected to be 10 and they can be negligible.

\begin{figure}
 \begin{center}
  \includegraphics[width=85mm]{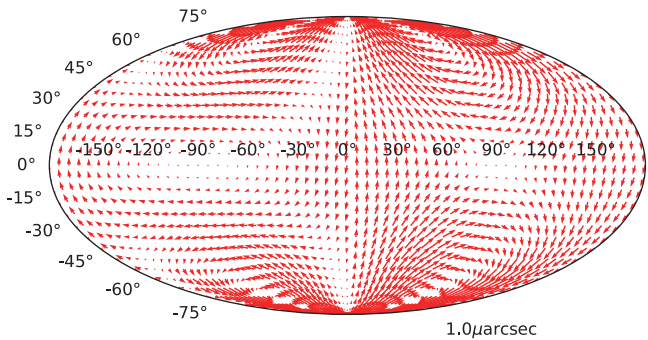}
 \end{center}
  \caption{Proper motions obtained from the {\it Gaia} data of the SDSS-{\it Gaia} QSO sample. The best-fit spherical harmonics for {\it Gaia} proper motion measurements are obtained by a calculation of the chi-square method. The red arrows show the vectors of proper motion fields. The length of reference for arrows ($1\,\mu $arcsec) is also shown at the right bottom.}  \label{fig.1}
\end{figure}

\section{Calculations and Results}\label{calculation_results}
	%最低限書かなければならないこと
		%①エネルギー密度$\Omega(f<10^{-9}Hz)$の上限値
	%②①の誤差

	%第1段落　-> ①エネルギー密度$\Omega(f<10^{-9}Hz)$の上限値
	%%第1文　図1に得られたproper motion fieldsのmapを示す。
	%%第2文　典型的なfieldsの強さは~0.1$\mu$ arcsecであった。
	%%第3文　このproper motion fieldに対応する重力波のエネルギー密度は
%%%%%　$\Omega_{\rm gw}<0.964\times 10^{-4}$である。
	%%第4文　これはD18の結果とconsistentであり、D18より制限が2桁向上した。
	%%第5文　この制限をstrain amplitudeに言い換えると$h_{\rm c}(f)<0.868\times 10^{-11}/(f/10^{-9}\, [{\rm Hz}])$となった。

We acquire proper motions with the SDSS-{\it Gaia} QSO sample, and perform fitting the proper motion fields with the spherical harmonics of equation (7) in Section \ref{Sec_method}. 
We obtain the best-fit spherical harmonics with the typical field strength of $\mathcal{O}(0.1)\, \mu$arcsec that is presented in Figure 1.
We estimate $\Omega_{\rm gw}(f)$ from the best-fit spherical harmonics with the equation \eqref{omegaGW}, and obtain the corresponding value  $\Omega_{\rm gw}(f) = 0.964\times 10^{-4}$. Because 
the 2 $\sigma$ deviation of the energy density of GWs is $3.804\times 10^{-4}$ (see Appendix B), %the upper limit becomes $\Omega_{\rm gw}(f) < 0.4768 \times 10^{-3}$ (95 \% CL).
the constraint becomes $\Omega_{\rm gw}=(0.964 \pm 3.804) \times 10^{-4}$ (95 \% confidence level (CL)).

This result is consistent with the one of D18 who use VLBI technique, while the sensitivity of the energy density of GWs has been significantly improved by more than one orders of magnitude from D18.
By using $\Omega_{\rm gw}(f)<0.4767\times 10^{-3}$ (95 \% CL) and equation \eqref{hc_OmegaGW},
we derive the upper limit on the strain amplitude 
$h_{\rm c}(f)<0.868\times 10^{-11}/(f/10^{-9}\, [{\rm Hz}])$.
%Omega and eq(XX) -> hc
%Constraint (<- what is the constraint??)  (method???) -> hc
 
\begin{figure*}
 \begin{center}
  \includegraphics[width=170mm]{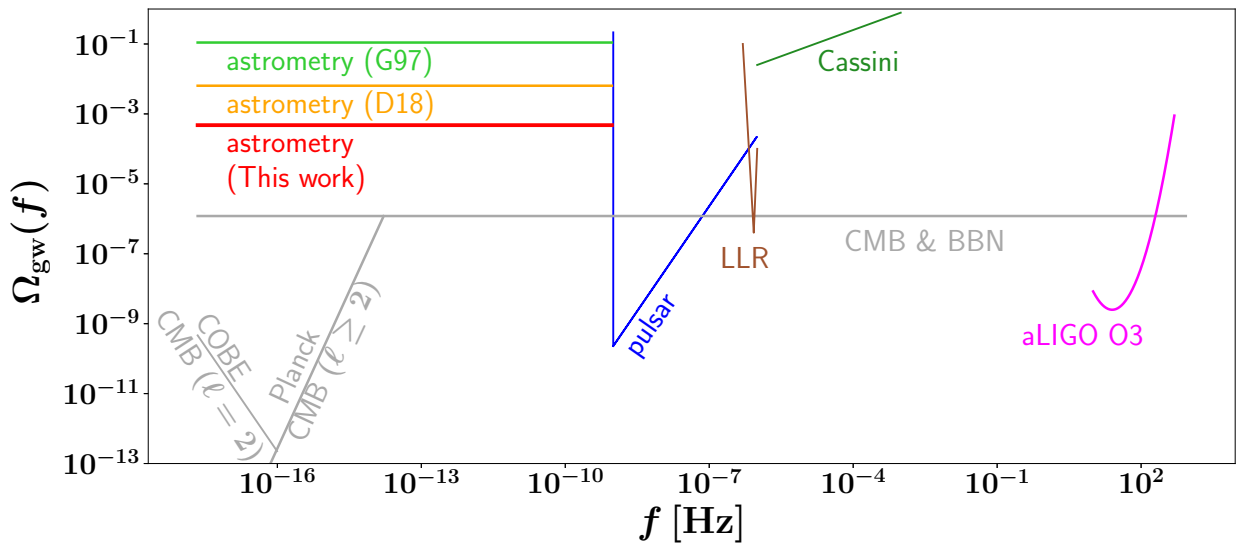}
 \end{center}
  \caption{{\it Gaia} constraint on the energy density of GWs shown with the red line. 
From low frequency to high frequency, the pink, dark green, brown, blue, 
green and orange lines represent 
the O3 run of the aLIGO (aLIGO O3) \cite{2018PhRvL.120i1101A},
Cassini \cite{2003ApJ...599..806A},
the Lunar laser ranging (LLR) \cite{1981ApJ...246..569M, 2021Univ....7...34B}, 
pulsar timing array (pulsar) \cite{2016PhRvX...6a1035L}, 
previous studies (G97, D18).
The gray lines represent constraints in the early Universe ($z>1090$) 
from CMB \& BBN \cite{2016PhLB..760..823P}, 
CMB at $2\le \ell \le 2000$ (Planck) \cite{2019PhRvD.100b1303N}, CMB at $\ell =2$ (COBE)\cite{2000PhR...331..283M}
%The line of CMB and BBN, Planck and COBE are constraints on the GWs only in the early Universe well before the cosmological recombination, that is, $z \gg 1090$. 
%過去の宇宙は全てグレーでそれ以外はカラー(今までの)
}  \label{fig.2}
\end{figure*}

\section{Discussion}\label{discussion}
	%最低限書かなければならないこと
		%①近傍宇宙のSMBH binary mergerが棄却されること
		%%第1文　SMBH(~10^{9}M_{\odot})の合体現象が発する重力波は10^{-9}Hz未満のもののみであり、LIGOなどの重力波観測装置では地面振動の理由で検出できない。
		%%第2文　SMBHが発する重力波の振幅はMatsubayashi et al. (2004)で次の式のように評価されている。
		%%第3文　この式に今回の結果を当てはめると10^{10} M_{\odot}の合体中のSMBH binaryは地球から80kpc圏内には存在しないことがわかる。

Super massive blackhole binary mergers emit GWs only at the frequency range $f \lesssim 10^{-9}\,[{\rm Hz}]$ because the size of the event horizon is very large. GWs at this range cannot be detected with laser interferometers on the ground such as aLIGO due to the large seismic noise in this frequency range. When we consider an inspiral phase of SMBH binary with the orbital radius $a$, we estimate the amplitude of GWs $h_{\rm expt}$ and the frequency $f_{\rm insp}$ in Matsubayashi et al. (2004) \cite{2004ApJ...614..864M}, hereafter M04, as 
\begin{eqnarray}
f_{\rm insp}&=&\dfrac{1}{\pi }\sqrt{\dfrac{GM_{\rm T} }{a^{3} } } \notag \\
&=&1.14\times 10^{-9}\times 
\left(\dfrac{a}{100R_{\rm grav}}\right)^{3\slash 2}
\left(\dfrac{M_{\rm T}}{10^{10}\,{\rm M}_{\odot}}\right)^{-1}\, ,
 \label{f_inspiral} \\
h_{\rm expt}&=&\sqrt{\dfrac{5}{32}}
\left( \dfrac{\pi^{2}G^{5}}{c^{12}} \right)^{1\slash 3}
\left( \dfrac{f^{2}}{M_{\rm T}} \right)^{1\slash 3}
\dfrac{GM_{1}M_{2}}{r} \label{h_expt} \\
&=&0.691\times 10^{-11}\times 
\left(\dfrac{M_{1}}{10^{10}\,{\rm M}_{\odot} }\right)
\left(\dfrac{M_{2}}{10^{10}\,{\rm M}_{\odot} }\right) \notag \\
&\times &\left(\dfrac{M_{\rm T}}{2\times 10^{10}\,M_{\odot}}\right)^{-1\slash 3}
\left(\dfrac{f}{10^{-9}\,[{\rm Hz}]}\right)^{2\slash 3}
\left(\dfrac{r}{400\,[{\rm kpc}]}\right)^{-1} \notag \, 
\end{eqnarray}
where $M_{1}$ and $M_{2}$ are the masses of the primary and secondary SMBHs, respectively. $M_{\rm T}$ is the total mass of the system, which is $M_{\rm T}=M_{1}+M_{2}$.
$R_{\rm grav}$ is defined as $R_{\rm grav}=2GM_{\rm T}\slash c^{2}$. $r$ is the distance of the binary system from the Earth. 

We adopt the fiducial values of $M_{1}=M_{2}=10^{10}\, {\rm M}_{\odot}$, 
$a=100 R_{\rm grav}$ and $r= 400$ kpc, 
and find $h_{\rm expt}=0.691\times 10^{-11}$. Because this value of $h_{\rm expt}$ does not meet the upper limit of $h_{\rm c}(f)<0.868\times 10^{-11}/(f/10^{-9}\, [{\rm Hz}])$, we exclude the existence of such SMBH binary systems at $r < 400$ kpc from the Earth. 
One can exclude existence of SMBH binaries with $R= 100R_{\rm grav}$ and $M_{\rm T}=2\times 10^{10}\,M_{\odot}$ in the local group.

%%式の定義は1回だけでそれ以外はスのまま使う。
%%このままだと情報がsplitするように見える。
%%論理ワープなし

	%②Gaia EDR3の現在の系	統誤差の影響		
%%第1文　Gaia EDR3の系統誤差により今回の論文で示した制限は
%%先行研究で予想された制限より100倍悪い。
%%第2文　その系統誤差の大きさは25 \mu arcsec/yr at G=15 magである。
%%第3文　この系統誤差は複数回の天体の観測と固有運動の観測モデルの
%%%%%　向上により今後のデータリリースで改善すると考えられている。
%%第4文　DR5ではこの系統誤差は5 \mu arcsec/yrとなり、それが達成されれば、
%%%　　　感度がエネルギー密度にして100倍向上し、先行研究で理論的に
%%%予言された感度や制限が達成される。
Our constraint on low frequency GWs is about 100 times weaker than a theoretical forecast (Book \& Flanagan 2011 \cite{2011PhRvD..83b4024B}). This difference between our constraint and the expectation is probably produced by the fact that the theoretical forecast only does not include systematic errors, but statistical errors. In the {\it Gaia} EDR3 data, the systematic errors become typically 400 $\mu $arcsec/yr at $G=20$, which is 10 times larger than the designed value of {\it Gaia}. One can expect that the systematic errors of {\it Gaia} will be reduced by improvements from the new multiple visit data and the proper motion model. According to the European Space Agency (ESA), the uncertainty on future releases of {\it Gaia} DR5 is expected to be suppressed to 40 $\mu $arcsec/yr at $G=20$. By this improvement, in DR5, one can constrain the existence of low frequency GWs with $\Omega_{\rm gw}\simeq 10^{-6}$ as Book and Flanagan (2011) \cite{2011PhRvD..83b4024B} predict.

%他のバンドとの比較、
%③重力波には感度が弱い振動数領域がある。
%%第1文　図2に現在の重力波の検出感度と制限を示して
%%第2文　観測感度は振動数の高い順に次の研究から得られている。
%%第3文　観測感度は10^{-4}<f<10^{0} Hzの領域は悪い。
%%第4文　観測感度が弱いのは地上の観測装置は地面振動により
%%　　　  低振動数の領域が見られないからである。
%%第5文　地面振動を回避する方法として人工衛星や人工天体が用いられてきた。
%%第6文　しかし、人工衛星や人工天体は
%% 	　　 電子回路のノイズ、アナログ–デジタル変換器の
%%　　　  規格の限界で観測感度に限界がある。
%%第7文　しかし、SMBH形成過程理解に必要なIMBH binaryはこの振動数領域の重力波しか出さないので感度の向上は必要である。
%%第8文　よってこの領域を観測するLISAやDECIGOなどのspace-based gravitational observatories が
%%　　　  この領域を観測するには必要である。
%%第7文　これらによりintermediate massive blackhole binaryの発見や、
%%　　　  インフレーション起源の重力波の観測が期待されている。
	%④電波干渉系VLBIの結果との比較, SKAについて
%(山内さんと議論して決める。)
We summarize the sensitivity of detection and upper limits of GWs 
in Figure 2. 
Figure 2 presents, from the high to low frequencies, aLIGO \cite{2018PhRvL.120i1101A}, 
a planetary exploration spacecraft (Cassini \cite{2003ApJ...599..806A})
the lunar laser ranging (LLR) \cite{1981ApJ...246..569M, 2021Univ....7...34B}, the pulsar timing \cite{2016PhRvX...6a1035L}, astrometry of the extragalactic objects (D18, G97 and this work), the abundance of relativistic components which affect the anisotropy of the cosmic microwave background (CMB) and the Big Bang nucleosynthesis (BBN) \cite{2016PhLB..760..823P}, the quadrupole temperature fluctuation ($\ell=2$) of CMB measured by {\it COBE} (COBE) \cite{2000PhR...331..283M}, and the CMB angular spectra ($\ell \ge 2$) (Planck) \cite{2019PhRvD.100b1303N}. Although Figure 2 does not include relatively high upper limits above $(\Omega_{\rm gw}(f) \gtrsim 10^{0})$, there exist constraints from a torsion bar detector (TOBA) \cite{2011PhRvL.106p1101I, 2018arXiv181201835S}, the global positioning system\cite{2014PhRvD..89f7101A}, seismic measurements of the Earth \cite{2014PhRvL.112j1102C}, a Lunar orbiter \cite{2014PhRvD..89f7101A} and another planetary exploration spacecraft (ULYSSESS \cite{1995A&A...296...13B}).

As for the future prospects, we expect to obtain the constraint of the energy density of GWs down to $(\Omega_{\rm gw}\sim 10^{-6})$ with the forthcoming {\it Gaia} DR5 data as discussed above in this section. This constraint is comparable with the one given by CMB \& BBN at $z\gg 1090$, but will be the strongest constraint at $z=0$. Moreover, this constraint is also comparable with the one for this low frequency range of $f[{\rm Hz}] \lesssim 10^{-9}$ expected by the experiment of the Square Kilometre Array (SKA) that will observe $10^{6}$ QSOs every month with the precision $\sim 10\,\mu$arcsec in the VLBI mode. In this way, the DR5 data of the {\it Gaia} astrometric satellite will provide the promising constraint at the low fraquency range. With this constraint, one can probe for mergers of SMBH binaries at a distance up to about 4 Mpc, and test the existence of SMBH mergers in the local group.

\section{Conclusion}\label{conclusion}
%最低限書かなければならないこと
	%第１段落　得られた重力波の上限値とその誤差→近傍宇宙のSMBH binary mergerが棄却されたこと
	%第２段落　望遠鏡が重力波の観測に有効であることが示されたこと。

%%第1文　本論文では長周期重力波に対して$\Omega_{\rm gw}<0.964\times 10^{-4}$という
%%		制限を可視光位置天文衛星Gaiaを用いて加えた。
%%第2文　この制限を得るために私たちはGaiaと分光同定されたQSOを持つ、SDSS QSOとのクロスマッチを行ってSDSS-Gaia sampleサンプルを作った。
%%第3文　Matsubayashi et al. (2004) のformulaを使うと
In this paper, we focus on the property of low frequency GWs, which creates apparent proper motions of astronomical objects. Because intrinsic proper motions of QSOs are negligible, we use QSOs for constraining the energy density of low frequency ($\lesssim 10^{-9}\,{\rm Hz}=(30\,{\rm yr})^{-1}$) GWs. We construct the Gaia-SDSS QSO sample with SDSS and {\it Gaia} data, cross-matching {\it Gaia} EDR3 sources and the SDSS QSOs, where the SDSS QSOs are spectroscopically confirmed. The Gaia-SDSS sample consists of 400,894 with the negligibly small number of contaminating objects only up to $\sim 10$.  

% abstractにある中間productのconstraintを説明する。->Omega_{\rm GW} %
We use the {\it Gaia} proper motion measurements for the QSOs in the Gaia-SDSS sample. While the astrometric measurement of {\it Gaia} is comparable to the one of VLBI, the number of QSOs is $\sim 500$ times larger than that of the VLBI study (D18). We obtain the best-fit spherical harmonics of proper motion of QSOs with the typical field strength of $\mathcal{O}(0.1)\,\mu$arcsec. 
On the basis of the relation between the proper motion and the energy density of GWs that is described in the equation \eqref{omegaGW}, 
%we obtain the upper limit of low frequency GWs 
we obtain the constraint of low frequency GWs 
($f<10^{-9}\,[{\rm Hz}]$) that is 
$\Omega_{\rm gw}=(0.964 \pm 3.804)\times 10^{-4}$.
By using formulae in M04, we exclude the existence of a SMBH binary 
within 400 kpc from the Earth including the Milky Way center and the local group with the fiducial setup.
%including nearby the Earth and our MW center with our fiducial setup. 

\section{Acknowledgements}
Numerical computations were carried out on analysis servers and Cray XC50 at the Center for Computational Astrophysics (CfCA), National Astronomical Observatory of Japan, Cray XC40 at the Yukawa Institute Computer Facility in Kyoto University. SA and MS acknowledges the Center for Computational Astrophysics, National Astronomical Observatory of Japan, for providing the computing resources of analysis servers and Cray XC50. 
This work was supported by the joint research program of the
Institute for Cosmic Ray Research (ICRR), University of Tokyo.
This paper is supported by World Premier International
Research Center Initiative (WPI Initiative), MEXT, Japan, and
KAKENHI (20H00180) Grant-in-Aid for Scientific Research (A) 
through Japan Society for the Promotion of Science.
MS is supported by JSPS KAKENHI Grant Nos. JP19K14718 and JP20H05859. DY is supported by JSPS KAKENHI Grant Nos. 17K14304, 19H01891.

\section{Appendix}
\subsection{Mathematical formulae of spherical harmonics}  
In this section, we show the mathematical formulae of eigenvectors of spherical harmonics. One writes the elements of the eigenvectors 
of spherical harmonics,  
$\vec{e}_{\theta} \cdot \vec{Y}^{\rm E}_{\ell m}$, 
$\vec{e}_{\varphi}\cdot \vec{Y}^{\rm E}_{\ell m}$, 
$\vec{e}_{\theta} \cdot \vec{Y}^{\rm B}_{\ell m}$ and 
$\vec{e}_{\varphi}\cdot \vec{Y}^{\rm B}_{\ell m}$ as follows. 
In the case that $|m| < \ell$, one has
\begin{eqnarray}
\vec{e}_{\theta}\cdot \vec{Y}^{\rm E}_{\ell, m}(\theta, \varphi)&=&
-\dfrac{1}{2}\left[\sqrt{(\ell - m)(\ell + m +1 )}e^{-i\varphi}
Y_{\ell, m+1}(\theta, \varphi) \right. \notag \\
&-&\left. \sqrt{(\ell + m)(\ell - m +1 )}e^{i\varphi}
Y_{\ell, m-1}(\theta, \varphi) \right]\, , \\
\vec{e}_{\varphi}\cdot \vec{Y}^{\rm E}_{\ell, m}(\theta, \varphi)&=&
-\dfrac{im}{\sin \theta}Y_{\ell, m}(\theta, \varphi)\, ,
\end{eqnarray}

\begin{eqnarray}
\vec{e}_{\theta}\cdot \vec{Y}^{\rm B}_{\ell, m}(\theta, \varphi)&=&
\dfrac{im}{\sin \theta}Y_{\ell, m}(\theta, \varphi)\, ,\\
\vec{e}_{\varphi}\cdot \vec{Y}^{\rm B}_{\ell, m}(\theta, \varphi)&=&
-\dfrac{1}{2}\left[\sqrt{(\ell - m)(\ell + m +1 )}e^{-i\varphi}
Y_{\ell, m+1}(\theta, \varphi) \right. \notag \\
&-&\left. \sqrt{(\ell + m)(\ell - m +1 )}e^{i\varphi}
Y_{\ell, m-1}(\theta, \varphi) \right]\, .
\end{eqnarray}

In the case that $m = \ell$, one has
\begin{eqnarray}
\vec{e}_{\theta}\cdot \vec{Y}^{\rm E}_{\ell, m}(\theta, \varphi)&=&
\dfrac{1}{2}\left[ \sqrt{(\ell + m)(\ell - m +1 )}e^{i\varphi}
Y_{\ell, m-1}(\theta, \varphi) \right]\, , \notag \\
& & \\
\vec{e}_{\varphi}\cdot \vec{Y}^{\rm E}_{\ell, m}(\theta, \varphi)&=&
-\dfrac{im}{\sin \theta}Y_{\ell, m}(\theta, \varphi)\, ,\\
\vec{e}_{\theta}\cdot \vec{Y}^{\rm B}_{\ell, m}(\theta, \varphi)&=&
\dfrac{im}{\sin \theta}Y_{\ell, m}(\theta, \varphi)\, ,\\
\vec{e}_{\varphi}\cdot \vec{Y}^{\rm B}_{\ell, m}(\theta, \varphi)&=&
\dfrac{1}{2}\left[ \sqrt{(\ell + m)(\ell - m +1 )}e^{i\varphi}
Y_{\ell, m-1}(\theta, \varphi) \right]\, \notag \\
& &.
\end{eqnarray}

In the case that $m=-\ell$, one has

\begin{eqnarray}
\vec{e}_{\theta}\cdot \vec{Y}^{\rm E}_{\ell, m}(\theta, \varphi)&=&
-\dfrac{1}{2}\left[\sqrt{(\ell - m)(\ell + m +1 )}e^{-i\varphi}
Y_{\ell, m+1}(\theta, \varphi) \right]\, \notag , \\
& & \\
\vec{e}_{\varphi}\cdot \vec{Y}^{\rm E}_{\ell, m}(\theta, \varphi)&=&
-\dfrac{im}{\sin \theta}Y_{\ell, m}(\theta, \varphi)\, ,
\end{eqnarray}

\begin{eqnarray}
\vec{e}_{\theta}\cdot \vec{Y}^{\rm B}_{\ell, m}(\theta, \varphi)&=&
\dfrac{im}{\sin \theta}Y_{\ell, m}(\theta, \varphi)\, ,\\
\vec{e}_{\varphi}\cdot \vec{Y}^{\rm B}_{\ell, m}(\theta, \varphi)&=&
-\dfrac{1}{2}\left[\sqrt{(\ell - m)(\ell + m +1 )}e^{-i\varphi}
Y_{\ell, m+1}(\theta, \varphi) \right]\, , \notag \\
& & 
\end{eqnarray}
where $Y_{\ell, m}$ is spherical harmonics, 
which is the solution of Laplace's equation.
$\vec{Y}^{\rm E}_{\ell, m}$ and $\vec{Y}^{\rm B}_{\ell, m}$ yield
\begin{eqnarray}
\int_{0}^{2\pi}\int_{0}^{\pi}
\vec{Y}^{\rm E}_{\ell               m}(\theta, \varphi)\cdot
\vec{Y}^{\rm E}_{\ell^{\prime} m^{\prime}}(\theta, \varphi)
\sin \theta d\theta d\varphi 
&=&\delta(\ell, \ell^{\prime})\delta(m, m^{\prime})~,\\
\int_{0}^{2\pi}\int_{0}^{\pi}
\vec{Y}^{\rm B}_{\ell               m}(\theta, \varphi)\cdot
\vec{Y}^{\rm B}_{\ell^{\prime} m^{\prime}}(\theta, \varphi)
\sin \theta d\theta d\varphi 
&=&\delta(\ell, \ell^{\prime})\delta(m, m^{\prime})~,\\
\int_{0}^{2\pi}\int_{0}^{\pi}
\vec{Y}^{\rm E}_{\ell               m}(\theta, \varphi)\cdot 
\vec{Y}^{\rm B}_{\ell^{\prime} m^{\prime}}(\theta, \varphi)
\sin \theta d\theta d\varphi 
&=&0 
(\forall \ell, \ell^{\prime}, m, m^{\prime} \in \mathbb{Z})~,\notag\\
& &
\end{eqnarray}
where $\delta (i,j) $ is the Kronecker delta such as 
\begin{equation}
\delta(i,j)=
\begin{cases}
1\,\,(i=j)\\
0\,\,(i\ne j)~.
\end{cases}
\end{equation}

\subsection{method of an error estimation}
In this section, we explain the estimation method of the error of the energy density of the long period GWs. The error is estimated by Fisher matrix method. In this paper, we define the likelihood function $\mathcal{L}$ as
\begin{eqnarray}
-2\ln(\mathcal{L})&=&\displaystyle\sum_{k=1}^{N_{\rm sample}}
\left\{ 
\dfrac{
\left(
\mu^{\rm obs}_{\theta, k}-\mu^{\rm th}_{\theta}
\right)^{2}
}{ 
(\sigma_{\theta, k}^{\rm obs} )^{2} 
} 
+
\dfrac{
\left(
\mu^{\rm obs}_{\varphi, k}-\mu^{\rm th}_{\varphi}
\right)^{2}
}{ 
(\sigma_{\varphi, k}^{\rm obs} )^{2} 
}
\right\}~.\notag \\
& &
\end{eqnarray}

The element of Fisher matrix $F$ is
\begin{eqnarray}
F_{X_{i}X_{j}} &\equiv &\dfrac{d^{2}}{dX_{i}dX_{j}}(-2\ln(\mathcal{L}))\notag \\
&=&2\displaystyle\sum_{k=1}^{N_{\rm sample}}
\left[ \dfrac{1}{(\sigma_{\theta,k}^{\rm obs})^{2}}
\dfrac{d\mu_{\theta}^{\rm th}(\theta_{k}, \varphi_{k})}{dX_{i}}
\dfrac{d\mu_{\theta}^{\rm th}(\theta_{k}, \varphi_{k})}{dX_{j}}\right.
\notag \\
& &+\left. \dfrac{1}{(\sigma_{\varphi,k}^{\rm obs})^{2}}
\dfrac{d\mu_{\varphi}^{\rm th}(\theta_{k}, \varphi_{k})}{dX_{i}}
\dfrac{d\mu_{\varphi}^{\rm th}(\theta_{k}, \varphi_{k})}{dX_{j}}\right]~.
\end{eqnarray}

One sigma deviation of a quantity $X_{i}$, $\sigma(X_{i})$, can be estimated with 
the inverse matrix of $F$, $F^{-1}$, as
\begin{equation}
    \sigma(X_{i})=\sqrt{(F^{-1})_{X_{i}X_{i}}}~.
\end{equation}
One sigma deviation of $\Omega_{\rm gw}$, $\sigma(\Omega_{\rm gw})$, is
\begin{eqnarray}
    \sigma(\Omega_{\rm gw}) &=& \dfrac{3}{2\pi \tilde{H}_{0}^{2}}
    \left\{
    \left[(E_{2, 0})\sigma(E_{2, 0})\right]^{2} \right. 
    +\left[(B_{2, 0})\sigma(E_{2, 0})\right]^{2} \notag \\
    & &+ \displaystyle\sum_{m=1}^{2}\left(
     \left[2{\rm Re}(E_{2, m})\sigma(E_{2, m})\right]^{2} \right. \notag \\
    & &+\left[{\rm Im}(E_{2, m})\sigma(E_{2, m})\right]^{2} \notag \\
    & &+\left[{\rm Re}(B_{2, m})\sigma(B_{2, m})\right]^{2} \notag \\
    & &+\left[{\rm Im}(B_{2, m})\sigma(B_{2, m})\right]^{2}
    {\Large {\text )}}{\Large {\text \}}}^{1\slash 2} ~.
\end{eqnarray}
In order to obtain 95 \% CL of the energy density of GWs, 
we show $\Omega_{\rm gw}\pm 2\sigma(\Omega_{\rm gw})$ as the constraint.

\bibliography{reference2021a}
\end{document}